# A systematic review and analysis of the viability of virtual reality (VR) in construction work and education

**Zia Ud Din[a,b], Payam Mohammadi[c], Rachael Sherman[c,d],**

[a,b]Department of Construction Management, University of Houston, 4730 Calhoun Road #300 Houston, TX 77204, USA. Email: uziauddin@uh.edu (Corresponding Author)

[c] Department of Engineering Technology and Construction Management, University of North Carolina at Charlotte, NC, USA

[d] Opower, Arlington, Virginia, United States

# **A B S T R A C T**

This systematic review explores the viability of virtual reality (VR) technologies for enhancing learning outcomes and operational efficiency within the construction industry. This study evaluates the current integration of VR in construction education and practice. Employing the Preferred Reporting Items for Systematic Reviews and Meta-Analyses guidelines, this review analyzed 36 peer-reviewed journal articles from databases such as the Web of Science, ERIC, and Scopus. The methodology focused on identifying, appraising, and synthesizing all relevant studies to assess the effectiveness of VR applications in construction-related fields. This review highlights that VR significantly enhances learning by providing immersive interactive simulations that improve the understanding of every complex construction process, such as structural elements or tunnel-boring machine operations. This review contributes by systematically compiling and evaluating evidence on using VR in construction, which has seen a limited comprehensive analysis. It provides practical examples of how VR can revolutionize education and work.

# 1. Introduction

The success of construction projects often depends on the effective use of tacit and explicit knowledge workers possess [1]. Tacit knowledge is usually acquired through hands-on practices such as internships or cooperative opportunities [2,3], whereas explicit knowledge is gained from conventional educational methods such as lectures, books, manuals, and databases [4]. However, delivering experiential knowledge in a traditional classroom setting is often challenging because of time and resource limitations [5].

In addition to the challenges of delivering hands-on experience in classroom settings, poor communication in the construction industry is a well-known challenge that affects construction project performance [6]. Information is predominantly delivered in 2D data formats, such as blueprints and specifications [7], which can be challenging for new workers to comprehend fully [8]. Consequently, this communication gap often results in decreased productivity, increased change orders, and compromised safety [9]. Adopting better communication methods that ensure timely and accurate information dissemination is crucial for enhancing overall construction project performance [10]. Implementing such an approach will address worker safety issues at construction sites, enhance collaboration among project stakeholders, and improve productivity.

Virtual Reality (VR) technology can address the challenges of providing hands-on experiences in classroom settings and improving the communication of design information in construction projects. VR offers immersive data visualization and interactive learning experiences, which allow students and workers to engage with complex concepts and scenarios in a safe and controlled environment [11,12]. VR can provide realistic training simulations, allowing workers to experience hazardous situations without associated risks and enhancing safety training effectiveness [13]. Additionally, VR improves collaboration among project stakeholders by offering a shared virtual space for real-time design reviews and modifications, thus reducing misunderstandings and errors [14]. Furthermore, VR enhances productivity by allowing workers to practice complex tasks in a virtual environment, leading to greater efficiency and accuracy when performing these tasks in reality [15].

Despite the recognized benefits of VR, its adoption in construction education and practice has been limited [16]. While recent studies have explored the potential of VR in construction education and construction projects, the adoption and depth of research are significantly lower than in other fields [16].

This gap suggests that the construction industry has not yet fully realized the potential benefits of VR technology.

To address this research gap, our study conducts a comprehensive systematic analysis to document VR's impact of VR on learning outcomes, safety training, and overall project management efficiency in the construction sector. We investigated (1) the impact of virtual reality on improving the quality of construction education and training (2) the evidence supporting increased VR usage and its impact on construction work quality, productivity, and safety. By providing a detailed review and meta-analysis, this study aimed to establish a clearer understanding of the potential of VR and guide future research.

## 2. Methodology

Systematic reviews follow a predefined search strategy to reduce bias in answering focused research questions by identifying, appraising, and synthesizing all relevant studies on a particular topic [17,18]. This systematic review was conducted in accordance with the Preferred Reporting Items for Systematic Reviews and Meta-Analyses (PRISMA) criteria. The database search utilized Web of Science, ERIC (EBSCOhost), and Scopus databases.

This section outlines the search parameters that identified 7,052 journal articles and describes how predefined exclusion criteria were applied to remove duplicate entries, publications outside the study's focus, and irrelevant studies, resulting in the inclusion of 36 papers. First, the authors searched a database using a predefined search string that included eight VR terms, six innovative education phrases, and 15 AEC terms, as shown in Table 1. The search results included journal papers with at least one virtual reality term, one innovative education term, and one AEC term in their titles, keywords, or abstracts. No term was eliminated to achieve the maximum width of the search results.

**Table 1** Systematic review search criteria.

| Category | Included terminology |
|---|---|
| Virtual Reality | ((virtual reality*) OR (virtual environment) OR (vr) OR ([v.r.]) OR ([head mounted device* VR]) OR (hmd*) OR ([hmd*]) OR ([Desktop-based] VR)) |
| Innovative Education | ((innovative learn*) OR (education*) OR (innovating teach*) OR (active*learn*) OR (interactive* learn*) OR (immersive* learn*)) |
| AEC | ((construction industry) OR (construction safe*) OR ([bim*]) OR (engineering design*) OR (structural design*) OR (civil engineering) OR ([building information model*]) OR (Architect*) OR (construction design*) OR (conceptual design*) OR (construct*) OR (build*) OR (structure* design*) OR (civil*)) |

The 7,052 records that met all the search criteria were transferred to the Covidence Systematic Review Software for article screening [19]. Before screening, 794 duplicate entries were identified and removed using the Covidence software. The authors then established exclusion criteria for a three-step screening procedure, which included (1) all non-construction applications, such as medical, dental, entertainment, and education; (2) all non-head-mounted device-based systems, such as computer monitors, TV screens, theaters, 3D TV, and 3D films; (3) all CAVE based systems; and (4) records in which VR was not the primary subject of the article. Two authors then performed a three-step screening procedure separately, and a third reviewer resolved any discrepancies. The first screening examined article titles and abstracts, eliminating 5972 records based on the predefined exclusion criteria. In the last phase, 286 full-text papers were evaluated, and 250 records were eliminated, leaving 36 deemed suitable for synthesis and inclusion in this systematic review.

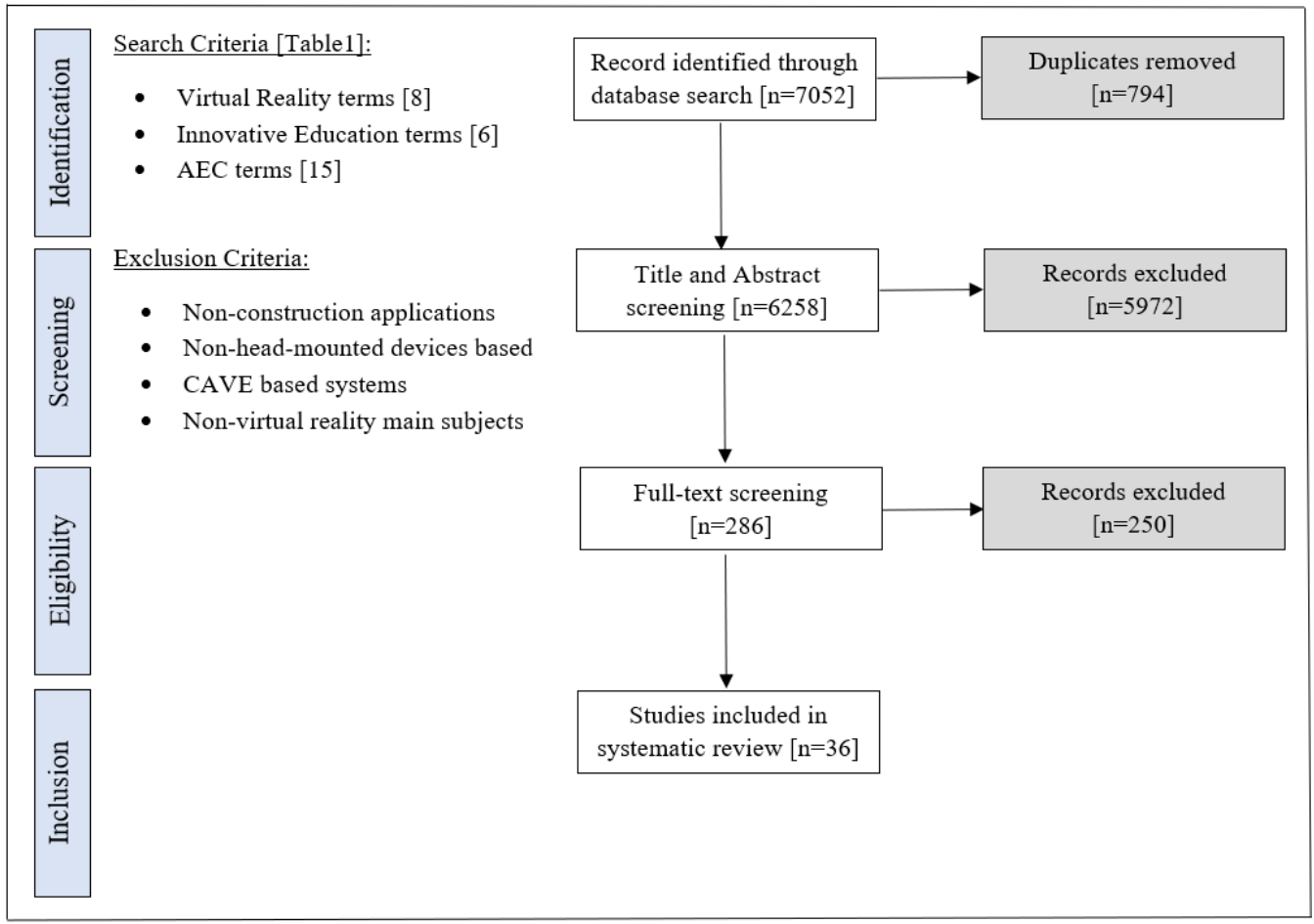

**Fig. 1.** The PRISMA flow diagram for this systematic review details the reduction in search results from 7,052 to 36 through three successive screening stages.

## 3. Literature Characterization and Bibliometric Analysis

This section describes the literature and bibliometric analysis of the 36 records included in this systematic review. First, a time-series analysis was used to identify the publishing of statistical patterns. Next, a location analysis determines the top-producing nations in virtual reality research. Journal analysis identified the most referenced publications and frequently represented journals. The concluding analysis explored the authors' co-occurrence to assess the frequency, relationship, and strength of these interactions.

*3.1*   Time Series Analysis

Fig. 2 shows the development of VR applications in construction research from 2011 to 2024 from the articles reviewed. This field's research grew modestly between 2017 and 2018, with a considerable increase between 2019 and 2020 and a sharp decline between 2021 and 2022. Before 2018, an average of 1 paper per year was published on VR. [20]The scope of research has expanded to include various aspects related to virtual reality (VR), such as applications, techniques, effectiveness analysis, logistics, cost, efficiency, influence on health and safety, architectural and structural design, and applications on construction sites.

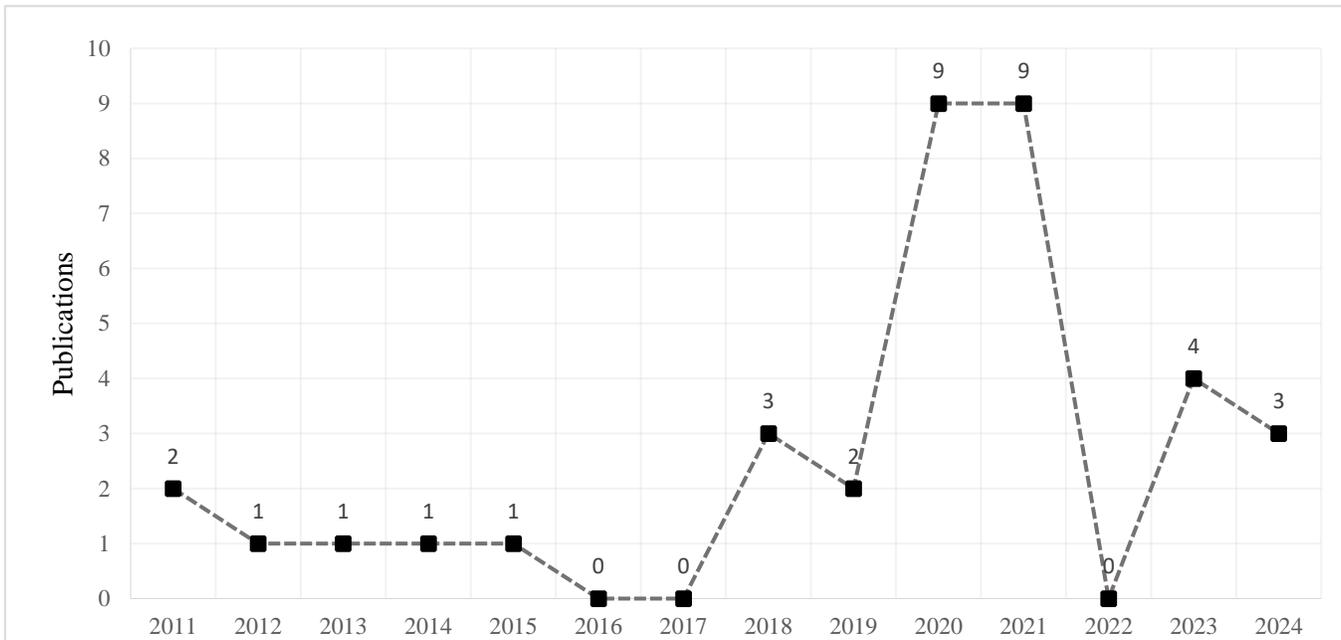

**Fig. 2.** Annual scientific production.

*3.2*   Location Analysis

Figs. 3 and 4 display the countries that produce the most articles and citations in the discipline. Twenty countries were included in the 36 articles included in the systematic review. The United States has led the world in total publications (12). China has four papers, South Korea and Indonesia each have two, and the following countries have one publication each: Australia, Canada, Egypt, Finland, Iran, Israel,

Japan, Kuwait, Pakistan, Peru, Portugal, Qatar, Thailand, Turkey, the United Arab Emirates, and the United Kingdom.

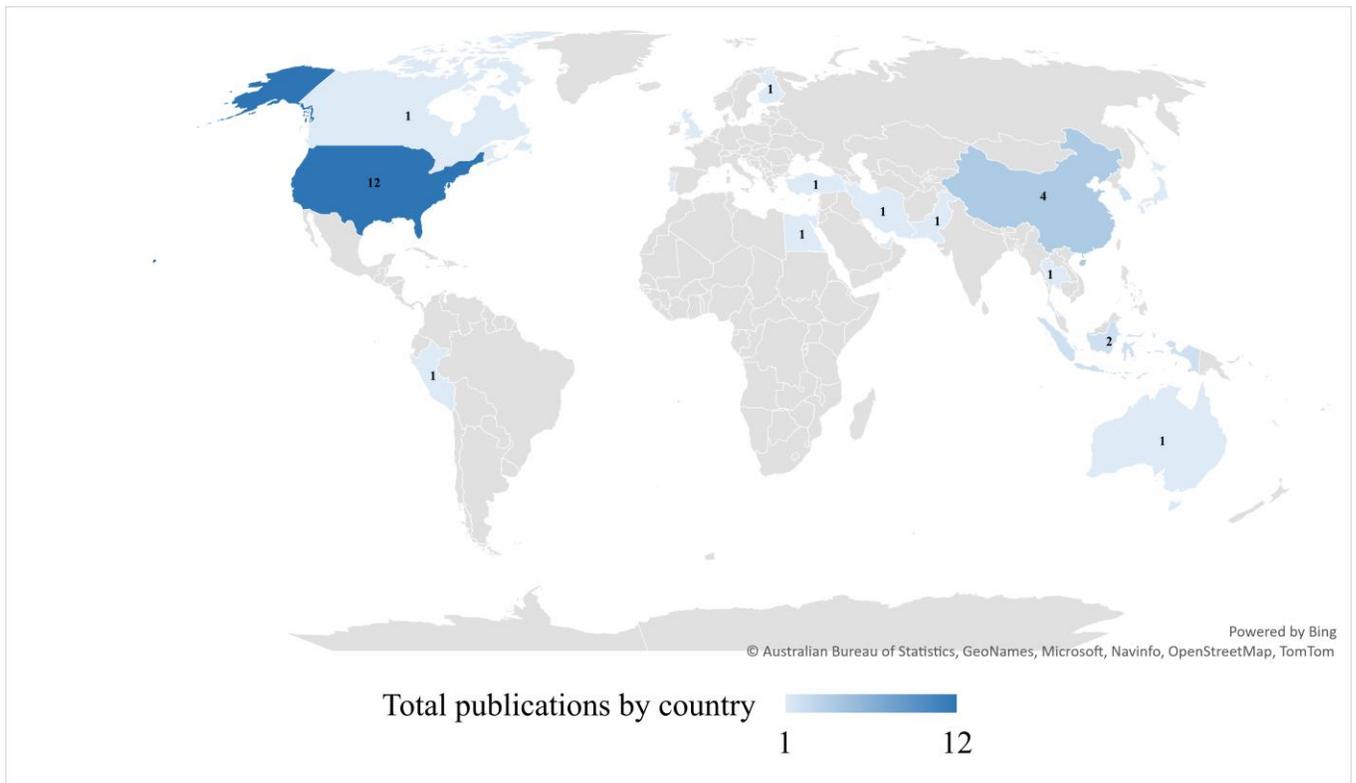

**Fig. 3.** Total publications by country.

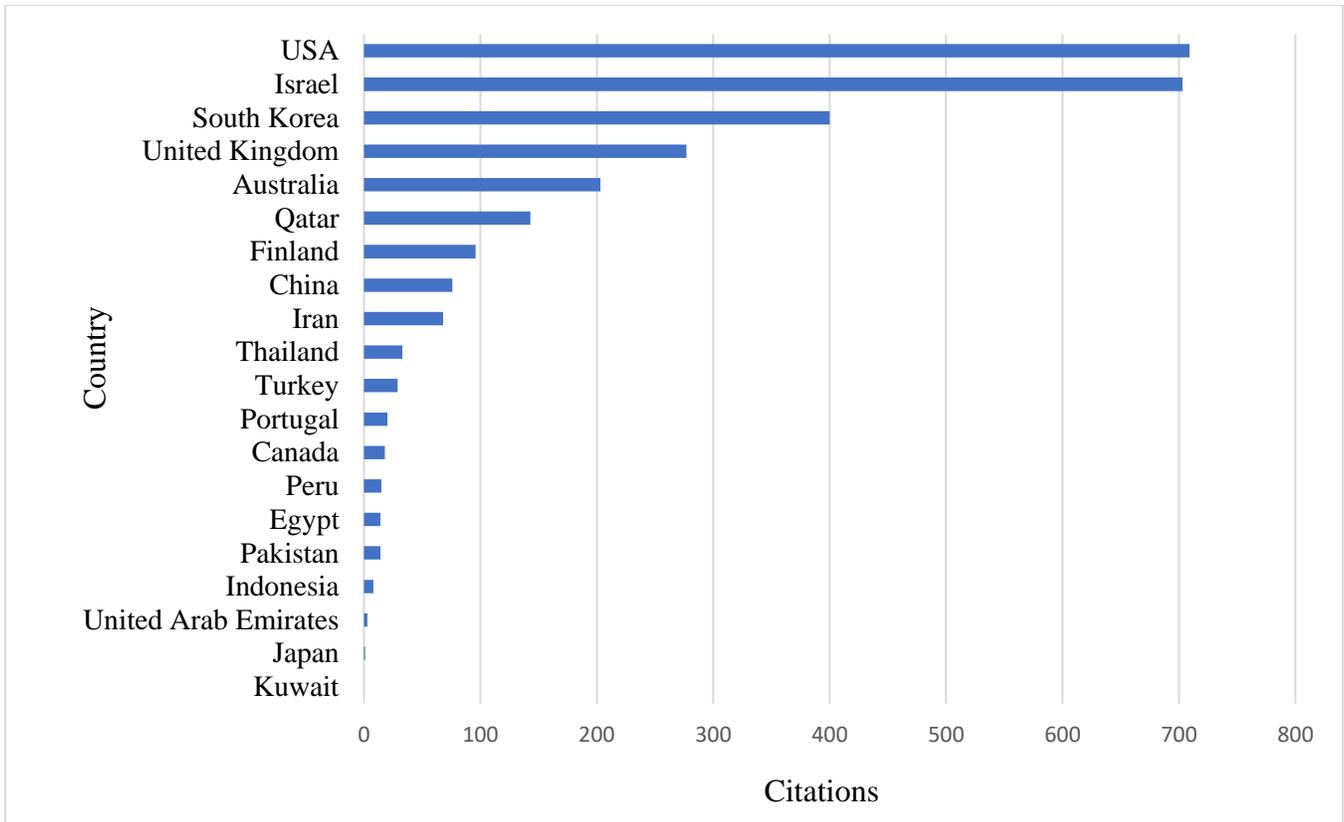

**Fig. 4.** Citations per country.

*3.3*  Publication Analysis

This study analyzed 36 journal articles. This research encompasses four construction-related fields: education (58.3% of publications), training (30.5%), construction management (5.5%), and design (5.5%). The International Council for Research and Innovation in Building and Construction (CIB) is the foremost publisher, with three journals and six publications. With 703 citations, Sacks et al. (2013) had the most significant number of citations. Table 2 lists the ten most cited papers.

Table 2. The most highly cited papers were included in this systematic review.

| No. | Source title, year | Title | Times Cited |
|---|---|---|---|
| 1 | Construction Management and Economics, 2013 | Construction safety training using immersive virtual reality | 703 |
| 2 | Journal of Intelligent & Robotics Systems, 2015 | A social virtual reality-based construction safety education system for experiential learning | 373 |
| 3 | Advanced Engineering Informatics, 2012 | Construction industry offsite production: a virtual reality interactive training environment prototype | 277 |
| 4 | Applied Sciences, 2020 | Digital twin and web-based virtual gaming technologies for online education: a case of construction management and engineering | 203 |
| 5 | Computers & Education, 2019 | Immersive virtual reality to enforce teaching in engineering education | 143 |
| 6 | Applied Ergonomics, 2021 | Implementing virtual reality technology for safety training in the precast/ prestressed concrete industry | 135 |
| 7 | Journal of Information Technology in Construction, 2011 | Building interactive modeling for construction education in virtual worlds | 119 |
| 8 | Engineering, Construction, and Architectural Management, 2020 | Development of virtual reality and stereo-panoramic environments for construction safety training | 116 |
| 9 | Journal of Safety Research, 2020 | Implementing and evaluating novel safety training methods for Construction Sector workers: results of a randomized controlled trial | 96 |
| 10 | Journal of Computing in Civil Engineering, 2020 | Evaluating the impact of virtual reality on design review meetings | 80 |

### 3.4 Author Analysis

Of the 36 publications in this systematic review, four authors were identified as having contributed at least two. As illustrated in Fig. 5, the VOSviewer software was utilized to create and visualize a co-authorship analysis of these four authors to determine the frequency, relationship, and strength of these co-occurrences. Co-authorship analysis has three essential components: circle size (frequency), colors (relationship), and lines (co-occurrences). Each circle symbolizes each author, and the size of the circle indicates the number of articles linked with the author's name, with a larger circle representing more publications. The larger the number of co-publications, the shorter the distance between authors. These co-publication relationships are defined by lines connecting the authors. The thicker the line, the more co-publications there are between authors. Finally, colors are used to represent the clusters of related authors. In this co-authorship analysis, the cluster sizes were set to be equal to two papers. Fig. 5 shows that Fadi Castronovo was the most productive author in this field [21,22]. In addition, Fadi Castronovo has the highest link strength. Two clusters represent groups of authors with at least two co-publications. There is strong cohesion between these groups but relatively little interaction between these clusters.

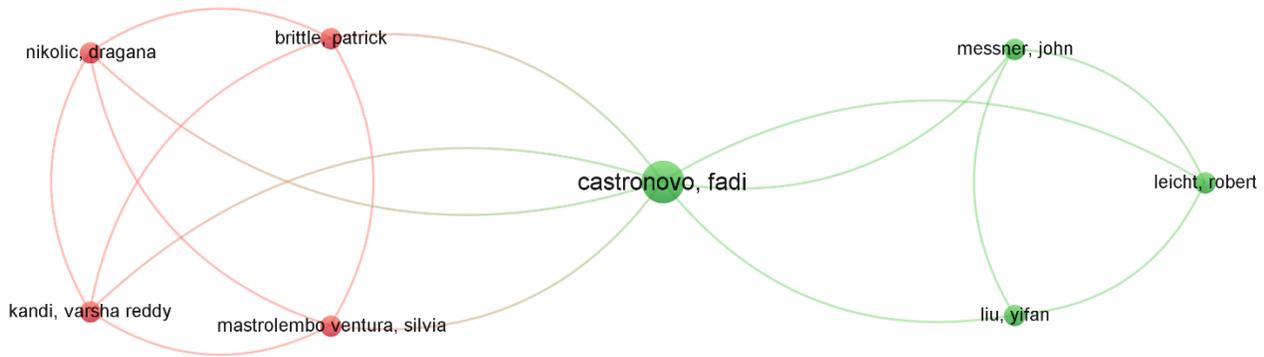

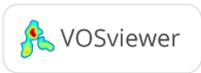

**Fig. 5.** Co-authorship network of authors with two or more publications.

## 4. Results: Evaluation of the Viability of VR in Construction Education and Training

This research aims to investigate the impact of VR on the AEC industry, focusing on its role in enhancing the quality of training, education, and professional work. Specifically, this research aims to evaluate the evidence supporting VR's increased use and its effectiveness in construction education and training. Additionally, it aimed to assess the impact of VR on the quality, productivity, and safety of construction work. Through this comprehensive evaluation, this study aims to provide a detailed understanding of how VR can be leveraged to improve various aspects of the AEC industry, thereby contributing to its overall advancement and innovation.

## *4.1. VR in Education and Training*

### *4.1.1 Enhancing Learning Outcomes and Engagement*

VR modules can enhance the learning experiences of AEC students [23–26]. Traditional teaching methods often use VR to provide immersive and interactive learning environments that simulate real-world construction processes, allowing students to gain practical experience and skills in virtual settings [23].

Halabi [27] discussed the use of VR in engineering education, highlighting its benefits in reducing development time and costs while increasing student motivation and creativity. They emphasized the effectiveness of VR and 3D prototyping in project-based learning (PBL) to promote effective communication, problem-solving skills, and enhanced learning outcomes.

Furthermore, Al-Khiami et al. [28] found that virtual reality (VR) significantly improved the motivation and performance of second-year civil engineering students in a concrete structures course when compared to traditional 2D drawings.

#### *4.1.1.1 Improved Understanding of Complex Concepts Through VR Simulations*

Through VR simulations, students can better comprehend complex concepts, such as the structural elements of a building or the operation of a tunnel-boring machine, as these concepts are often difficult to grasp using traditional teaching methods [23]. Such pedagogy requires developing a syllabus that combines lectures on theoretical concepts with VR tutorials, providing a hands-on experience through immersive learning [23]. Another example of VR for training is its use in operating unmanned aerial vehicles (UAVs) for building inspections. Researchers developed a VR-based flight training simulator, DroneSim, as an alternative to real-world drone-mediated building inspection training. The simulator enhanced construction students' UAV operational and flight training skills. The findings revealed that DroneSim, representing a virtual training group, acted as an efficient alternative to traditional real-world inspection training, offering comparable, if not superior, results [29].

Moreover, VR offers a more intuitive grasp of spatial dimensions and other model features than conventional learning materials, which rely on rendered images [30]. A study utilizing VR headsets to explore wood frame construction assemblies found that students understood the subject matter better than those using traditional resources [31]. This finding was further supported by the success of LADUVR, an educational VR application designed to learn architectural details. The step-by-step simulator for the house-building process within LADUVR effectively demonstrated the superiority of 3D real-scale presentations over traditional construction courses. While the initial feedback from participants highlighted some missing steps in the early sections of the simulator, they suggested that increasing the number of steps significantly enhanced their comprehension of the building process. Subsequent sessions received excellent ratings for the clarity of the data presented, emphasizing the value of detailed, immersive educational tools such as LADUVR [32].

Aydin and Aktas [33] discussed the integration of VR in architectural design education, highlighting the development of a novel curriculum based on student feedback. It explored two digital design ecosystems tested in undergraduate courses, emphasizing advanced digital design methods and techniques, such as 3D modeling, building information modeling, visual programming, and real-time rendering. They presented the effectiveness of these ecosystems through a User Experience Questionnaire, revealing that VR tools are attractive and stimulating but can be complicated and confusing. Their findings offer insights into designing VR design studios, addressing both task-related and non-task-related qualities.

*4.1.1.2  Bridging the Gap Between Theory and Practice*

VR helps bridge the gap between theory and practice by providing immersive environments that simulate real-world construction processes and enhance practical skills. VR environments can present detailed simulations of complex processes [34]. For example, Sepasgozar [23] provided comprehensive details for every construction task, offering twenty-minute modules that demonstrated the complete operation of a piling rig. This included the sequential processes of digging holes, inserting the cage, pouring in situ concrete, and capping off. Such detailed modules enable students to observe the sequences and predecessors of each method.

Dunston et al. [35] highlighted that VR-based training platforms, which include mock-ups of equipment controls and realistic 3D-modeled scenarios, can simulate various operational conditions, thus avoiding the hazards and costs associated with field training. Their findings indicate that while VR simulators effectively develop essential perceptual-motor skills required for operating complex machinery, assessing the true transfer of these skills to real-world tasks remains a significant challenge.

A VR interactive training environment offers a safe platform for mastering offsite production (OSP) techniques. This VR solution is designed for professionals, including architects, suppliers, manufacturers, and project managers, to promote multidisciplinary learning and break down knowledge silos [36].

By using virtual reality (VR) environments, students learn better about construction project management. In a study, educators observed a substantial improvement in students' comprehension of construction project management through the use of the Second Life (SL) platform. Participants in SL actively participated in project-based learning, tackling various challenges in construction project management from the initial design phase to project completion. This approach focused on aspects including resource allocation, cost management, quality assurance, and time efficiency [37].

The interactive nature of VR increases students' motivation and interest, making it engaging and beneficial to their learning experiences. Students participating in a study reported positive experiences with virtual modules, expressing increased motivation, driving knowledge, and applying learned concepts in real construction scenarios [23,38]. The virtual modules provide students with a deeper understanding of construction practices by simulating productivity measures, visualizing machinery operations, and illustrating critical paths in construction projects, leading to increased knowledge and skill acquisition [23]. VR in construction education can offer students a realistic and engaging learning environment that bridges the gap between theory and practice [39].

*4.1.2 Facilitating Collaborative Learning and Decision-Making*

VR facilitates collaborative learning and decision-making through interaction with the content. For example, a social, collaborative VR-based construction safety education system was developed using five criteria: ease of use, visual output, safety cognition, safety memory, and safety accessibility [15]. The system demonstrated significant potential for enhancing experiential learning in terms of construction safety. Another notable application of VR in construction education was the use of role-playing experiences in the virtual environment called SL to engage students in interactive scenarios for construction process learning, safety training, design reviews, and real-time integration of physical sensors for building monitoring [39]. However, using Virtual Reality (VR) in education poses several challenges, including high technical knowledge to design scenarios, a steep learning curve, high costs, realism and accuracy issues, difficulties integrating curriculum, user discomfort, and ethical and privacy concerns [37].

*4.1.3   VR Implementation and Perception in Construction Education*

VR can be experienced in several ways, including through headsets and computer screens [34]. Students positively view the implementation of head-mounted devices in the construction curriculum, indicating a favorable perception of its potential benefits [31]. To further facilitate the application of VR in construction education, researchers have proposed a framework that utilizes the Unity tool to create content for construction sequence simulations. This framework allows for rapid content creation [40].

Additionally, VR technology has proven cost-effective and accessible for educational purposes. The implementation of VR reduces the financial burden of deploying experiential learning platforms in universities, making it a cost-effective solution for academic purposes [41,42].

*4.1.3.1  Reduced Training Costs and Resources*

Despite the upfront investment required for application development, the cost of implementing VR-based applications is significantly lower compared to the expenses associated with real-life equivalents, such as organizing visits to actual construction sites [32]. In addition to cost savings, VR applications have the potential to shorten the educational process for intricate laboratory tasks that involve numerous steps, further optimizing both time and cost efficiency [43]. By streamlining the learning process and reducing the time required for students to master complex concepts and procedures, VR technology can help educational institutions maximize their resources and provide a more captivating and efficient educational experience. Moreover, compared to traditional paper-based testing, the temporal demands during short testing quizzes (STQs) utilizing VR technology are significantly lower [44].

*4.1.3.2  Scalability and Flexibility of VR Solutions*

The scalability and flexibility of VR solutions make them suitable for various educational settings, ranging from small classrooms to large training facilities [42]. In construction education, this approach allows students to visit construction sites virtually and navigate through various floors and stages of construction [42]. By providing a virtual alternative to physical site visits, VR technology enables students to access learning opportunities that may otherwise be limited owing to logistical, safety, or accessibility constraints. When an immersive interface offers virtual access to construction systems, it eliminates the need for physical site visits to observe these systems in their actual context [45].

In a study by Maghool et al. [29], almost all participants highlighted the superior flexibility of this VR application compared to traditional university classes. They appreciated the freedom to choose when, where, and what they had learned. The ease of shifting between different learning resources, from a three-dimensional representation of construction details to a recorded video demonstration of construction activity, was particularly noted for enhancing engagement. Additionally, approximately half of the participants believed that this flexibility level would significantly improve the effectiveness of their learning process [32]. Consequently, VR technology makes the learning experience more inclusive and accessible, promoting equal chances for students to employ the subject matter and expand their skills in construction education [41].

*4.1.3.3   Safety Training and Hazard Identification*

VR is instrumental in safety training and hazard identification in the construction industry [46]. When VR solutions implement social collaborative virtual reality, it significantly enhances safety education [15]. In this case, multi-user collaboration and role-playing functions facilitate a deeper understanding of construction site dynamics and hazards. Simultaneously, the safety content provided through the system improves safety cognition by offering adequate information via virtual scenarios and safety standard-based question-answer games, enabling students to acquire safety knowledge effectively [15].

VR can simulate real-world safety circumstances, allowing students to exercise and understand safety protocols in a controlled environment. According to participants in the study conducted by Jeelani et al. [47], VR technology training exposed them to various construction activities, hazards, and accidents that would not be feasible in a single accurate construction site. This VR training helped them comprehend how seemingly minor hazards could trigger a chain reaction, potentially resulting in serious accidents [47]. Moreover, such a module empowers students to understand construction site risks and potential dangers [15].

By utilizing VR technology to simulate safety hazards, it is possible to simulate safety hazards without using materials or damaging equipment [41,48]. In a study conducted by Jeelani et al. [47], it was observed that after training, participants recognized more hazards and employed more efficient strategies to manage them. Additionally, a study introduced a social collaborative VR-based system to enhance learning about construction safety by allowing students to engage in role-playing, collaborative learning, and social interaction within a 3D virtual environment [15].

Joshi et al. [49] discussed developing and implementing a VR training module for safety training in the precast/prestressed concrete industry. The VR module, developed using Unity3D and Visual Studio, focused on three major safety concerns: personal protective equipment (PPE), the stressing process, and suspended loads. It aimed to provide cost-effective, repeatable, and engaging safety training. The module was evaluated through efficacy and effectiveness analyses, showing it to be user-friendly with minimal simulation sickness and more effective than traditional video-based training methods in engaging participants and enhancing their understanding of safety protocols. These findings demonstrate the effectiveness of VR in improving the safety awareness and hazard identification skills of construction students and professionals.

*4.2* **VR in Construction Work**

*4.2.1   Facilitating Iterative Design Processes*

VR is a powerful tool for iterative design processes, facilitating efficient progression toward an optimal design solution. It simplifies "what-if" conversations with stakeholders, helps to visualize challenges, explores design options, and assesses the impact of various alternatives on aesthetics and cost [50].

Users can choose architectural components, modify materials, compare cost variations between designs, and see the effect on the overall project cost in real time [50]. This hands-on design exploration approach fosters creativity, encourages experimentation, and facilitates informed decision-making. Furthermore, VR's ability to present complex architectural concepts in an accessible and engaging manner enhances communication among project participants, streamlines the design process, and leads to more successful outcomes. Consequently, VR technologies are highly valued by students and professionals alike for their ability to visualize architectural models, making complex concepts more accessible and discussions productive [48]. This immersive technology enhances the design process and improves communication and decision-making among project participants.

*4.2.2 Enhancing Design Review and Error Detection*

VR significantly enhances the design review process by providing designers with a sense of presence in their virtual environment. This immersive experience allows users to perceive and interact with design elements more effectively and intuitively. It offers insights that are impossible to gain through traditional methods, which often fail to provide a clear spatial understanding, convey complex designs effectively, offer interactivity, reduce cognitive load, engage reviewers, review changes flexibly, and provide a holistic perspective [22].

VR improves spatial perception and visualization, enabling users to better understand and interact with 3D models. These models are particularly effective for examining specific design concerns, such as sightlines, owing to the first-person perspective and capability for free movement within the virtual space [22]. VR's high-quality visualization capabilities are especially beneficial for architectural designs that include complex features, such as curved walls, slanted ceilings, or other irregular shapes, allowing for a more accurate representation of unique architectural details [22].

In VR environments, users can experience being inside a virtual house, freely walking through it, and interacting with different building parts. By clicking on highlighted points, users can access detailed views of specific building sections through interactive menus and section planes [32]. This interaction allows users to look through parts of the building to comprehend architectural details [32].

Moreover, it helps identify design issues and model mistakes early in the project, reducing the likelihood of costly errors. For example, a game Design Review Simulation in VR significantly enhanced the ability of students to identify mistakes, with results showing that they detected a notably higher number of errors compared to reviewing 2D drawings [21]. Additionally, VR facilitates the discovery of design issues or modeling mistakes through the features of virtual teleportation, which offers an eye-level viewing perspective. This method provides a more detailed and intimate view of the virtual environment, allowing for a thorough examination of design elements that may be overlooked in traditional viewing modes [48]. Not only does design comprehension improve, but VR also allows users to identify errors, for example, in construction schedules [11].

*4.2.3   Transforming Collaboration and Decision-Making*

VR and Building Information Modeling (BIM) reshape the way abstract concepts are understood and applied across professional communities, encouraging cross-cultural and global interactions [37]. BIM with VR helps students understand design (for example, complex plumbing networks) more effectively [51]. This technology has proven exceptionally effective in interior design and construction, particularly in selecting and visualizing finishing materials. This assists various stakeholders in choosing their preferred finishing materials and visualizing design changes, thereby enhancing the decision-making process with recommendations on material collocation [50]. VR immerses users in a virtual environment, allowing them to experience designed spaces qualitatively and quantitatively. This immersive experience is pivotal in evaluating spatial designs and making informed decisions [50]. VR's facilities to present designs to clients make complex proposals more accessible and understandable [52].

BIM combined with VR technology enhances this process by enabling users, even those with minimal software skills, to actively engage in collective decision-making. Platforms such as InsiteVR enhance this capability by providing real-time dynamic design collaboration experiences, allowing people worldwide to meet and work virtually [48].

Through role-playing scenarios, participants can interact with building systems and construction equipment, along with other project stakeholders, throughout the various phases of a project, including programming, design, construction, and operations and maintenance. This interaction allows for efficient management, simulation, analysis, and updating of project data, facilitating a comprehensive and practical approach to resource allocation [39]. Integrating VR and BIM in construction management ultimately leads to more accurate planning, reduced waste, and improved project outcomes.

*4.2.3.1   Providing Realistic and Immersive Experiences*

VR provides realistic and immersive experiences invaluable for construction work. The immersive quality of VR technology offers an experience akin to visiting a real construction site where participants can observe work unfolding as if they were physically present. This realism is emphasized by users who describe their experiences as believable and engaging, and it feels like an educational excursion into the actual workings of a construction site [23,47]. Realistic 3D views provided by VR applications enhance this sensation, allowing users to see and feel the space as if it were real [43,52].

Furthermore, VR technology has transformed architectural design, which is presented and reviewed. It enables a real-time, photorealistic experience from a first-person perspective. It incorporates elements such as gravity to visualize designs within their true context, far surpassing traditional 2D drawings and static 3D models on a computer screen [48]. This technology also allows users within a VR interface to toggle the visibility of construction elements. This enables them to visualize building components typically obscured by other layers, thereby offering a deeper understanding of structural complexities [45]. Daylighting and artificial lighting simulations in VR, complete with immersive walk-through and fly-through experiences, are phenomenal. This technology offers a vivid and interactive way to explore and understand the effects of lighting in various settings, thereby enhancing both the aesthetic and functional aspects of architectural and design education [48].

*4.2.3.2 Enhancing Understanding of Scale and Complexity*

VR enhances the understanding of the scale and complexity of construction projects and prepares professionals for real-world challenges. Virtual reality (VR) has proven to be a superior method for hazard identification and prevention training, particularly for tasks such as reinforced concrete and stone cladding works, significantly outperforming traditional methods [14]. VR technology creates immersive and interactive scenarios for users to engage with during assessment tasks, such as site inspections and job safety analysis reviews. This approach enhances training by accommodating diverse contextual safety skill assessments within simulated environments, allowing for practical, experiential learning [44]. Learners are introduced to innovative approaches such as the safety pyramid, which shifts their focus from merely protecting against hazards with personal protective equipment (PPE) to actively eliminating them [47]. This proactive approach is reinforced through feedback mechanisms and virtual site walkthroughs that sharpen risk perception by highlighting potential dangers under seemingly safe conditions [47].

By engaging with various modules, trainees can visualize the scale of operations and possible impacts of risks, enhancing their understanding of the repercussions of safety failures in real-world settings [23]. In VR, students inspect and identify hazards within 3D models, reflecting on the safety information learned from previous modules [15]. This learning is further deepened as they participate in safety training games by applying their knowledge to recognize latent hazards during simulated construction activities [15].

Studies have shown that VR-based safety training enhances self-efficacy more effectively than traditional lecture-based training by providing mastery experience and interactive learning opportunities [53]. Additionally, virtual environments offer a safe alternative to actual construction sites, eliminating the inherent risks of physical site visits [32].

Detailed procedural training includes checks such as ensuring the correct wearing of PPE before entering hazardous areas such as wells, assessing mental readiness and equipment suitability, and familiarizing trainees with underground conditions and safety equipment such as refuge chambers under NPC guidance [54]. VR also simulates real-world contexts to review crucial safety issues, such as egress, clearance, and traffic maneuvering during various construction and maintenance phases [48].

Furthermore, VR facilitates interactive learning through quizzes and role-playing scenarios, allowing users to engage in hands-on activities with various safety aspects of construction equipment and operations, making the learning process both immersive and effective [39]. These features highlight the comprehensive nature of VR-based safety training, which addresses many safety concerns and provides learners with practical experience for identifying and mitigating hazards.

*4.2.4 Resource Allocation and Project Planning*

Allocating people and machinery to specific tasks is crucial for construction and project management [23]. VR technology aids in optimizing equipment and labor utilization by providing visualization of detailed simulations of project workflows. Virtual environments such as Second Life (SL) offer a range of transformative possibilities for education and professional collaboration. SL, for instance, enables group discussions with learners globally and facilitates the exploration of equipment and facility simulations, thus opening extensive learning and interactive opportunities [37].

By identifying potential issues early and allowing for detailed planning, VR minimizes rework and change orders, saving time and resources. Using VR in project planning can minimize sudden and time-consuming changes before construction begins, which enhances end-user satisfaction and reduces the costs related to changing orders during the construction phase [50].

## 5. Discussion

This systematic review aimed to evaluate the viability of VR in construction work and education. The findings from the reviewed studies demonstrate that VR has significant potential to enhance learning outcomes, improve safety training, and optimize resource allocation in the construction industry.

### 5.1 Enhanced Learning Outcomes and Engagement

VR significantly enhances learning outcomes by providing immersive and interactive simulations that improve understanding of complex construction activities. This is consistent with findings by Sepasgozar [23], who reported that VR enhances the learning experience for AEC students by combining traditional teaching methods with immersive learning environments. Similarly, Jeelani et al. [47] found that VR-based training improves students' comprehension of machinery mechanics, underscores the potential of VR in improving teaching effectiveness, and hence helps bridge the gap between theory and practice by providing hands-on experience in a virtual setting.

However, some studies, such as Maghool et al. [32], suggest that the initial implementation of VR might face challenges related to the completeness and detail of the simulations. Participants in their study indicated that early versions of VR applications were missing critical steps in construction processes, which were later improved. A detailed and comprehensive VR module is needed to maximize learning benefits.

In addition to the challenges related to the content of VR simulations, there are other factors to consider for successful implementation. In vocational education, Ravichandran and Mahapatra [55] highlight challenges such as the expenses for equipment and software and the requirement for specialized technical support. Similarly, Lie et al. [20] identify barriers such as expenses, technology, and the importance of careful design and evaluation for successful implementation in health professions education. While particular challenges have been examined within the context of specific educational domains, these findings may be transferable and applicable to studies in AEC when VR technologies are employed.

*5.2     Facilitating Collaborative Learning and Decision-Making*

The role of VR in enhancing collaborative learning and decision-making is well-documented in existing literature. Le et al. [15] demonstrated that a social, collaborative VR-based construction safety education system significantly enhances experiential learning. Other studies corroborate this, showing that VR facilitates real-time collaboration and communication, allowing students and professionals to work together more effectively. Ku [39] further supports this by illustrating how VR and BIM enable real-time collaboration for design reviews and facility management.

However, some researchers have pointed out that the time required to create VR scenarios can hinder widespread adoption [34,56,57]. While VR has great potential for collaborative learning, the development process must be streamlined to increase its practical usability.

*5.3     VR Implementation and Perception in Construction Education*

The positive perception of VR in construction education is echoed across multiple studies. Lucas [31] found that students viewed the implementation of VR headsets favorably, which aligns with our findings that VR technology is well-received by students for its potential to enhance learning and engagement. Also, Walker et al. [42] highlighted the cost-effectiveness and scalability of VR solutions, which we identified as significant advantages of VR in education. This is corroborated by research in which researchers developed a toolkit to help overcome cost-related hurdles and manage VR implementation for many students, making VR a scalable option for distance learning. However, our review also noted the initial high costs and resource requirements for VR application development, a challenge similarly identified by Maghool et al. [32]. This suggests that while VR offers long-term benefits in cost reduction and resource optimization, the initial investment remains a critical consideration for educational institutions.

*5.4     Safety Training and Hazard Identification*

Our review found that VR significantly improves safety training and hazard identification, a finding supported by several studies. Jeelani et al. [47] demonstrated that VR-based safety training exposes trainees to various hazards in a controlled environment, enhancing their ability to identify and manage risks. Le et al. [15] also showed that VR improves safety cognition through interactive role-playing and collaborative learning experiences. However, some studies, such as those by Nykänen et al.

[53], suggest that while VR enhances self-efficacy in safety training, its effectiveness compared to traditional methods may vary based on the specific training context. Hence, VR training modules must be tailored to particular construction tasks to maximize their effectiveness [58].

*5.5  Transforming Collaboration and Decision-Making in Construction Work*

Our findings indicate that VR transforms collaboration and decision-making in construction work by providing realistic and immersive experiences. This aligns with Balali et al. [50], who found that VR simplifies design review processes and enhances stakeholder communication. Liu et al. [22] further support this by demonstrating that VR improves spatial perception and visualization, making complex design elements more accessible.

However, the effectiveness of VR in identifying design issues early in the project lifecycle, as noted in our review, suggests a need for ongoing evaluation and refinement of VR tools to ensure they meet the evolving needs of the construction industry. This aligns with the findings by Alizadehsalehi et al. [48], who emphasized the importance of continuous improvement in VR applications for construction.

**Future Directions**

The authors believe, in the future, based on this systematic review, that conducting longitudinal studies to assess the long-term impact of VR training on both student learning and career outcomes for construction professionals would be beneficial. These studies would help determine the sustainability and transferability of skills acquired through VR to real-world settings. Research should also explore developing cost-effective VR solutions to democratize access, particularly in lower-resource settings, and investigate the integration of VR with other technologies, such as augmented reality (AR) and mixed reality (MR), to create more comprehensive educational tools.

Our systematic review documents the significant benefits and acknowledges the challenges of VR in construction work and education. VR's enhanced learning outcomes, collaborative opportunities, and safety training benefits are well-supported, while initial costs and development time remain relevant challenges. While the use of VR has been explored for various potential benefits in construction education and professional work, and multiple studies have corroborated its advantages, future research should focus on addressing the identified challenges to fully realize the transformative potential of VR in the construction industry.

Best practices for implementation and cost-benefit analysis need to be thoroughly documented and addressed to move beyond the testing phase and ensure sustainable implementation. By doing this, the authors believe VR can become integral to construction education and practice, improving quality, safety, and productivity in the sector.

**Implications**

Adopting VR technology in construction education can revolutionize how future professionals are trained, providing immersive and interactive learning environments that enhance the comprehension of complex concepts and improve practical skills. This can lead to a more skilled and safety-conscious workforce, reducing workplace accidents and increasing productivity. Construction firms would benefit from integrating VR into their operations to streamline the design processes, improve stakeholder collaboration, and enable more efficient project planning and execution. Following these processes and utilizing VR as a tool can result in cost savings, reduced project timelines, and higher-quality outcomes.

Moreover, the scalability and flexibility of VR solutions make them accessible to a broader range of educational settings, from small classrooms to extensive training facilities, promoting inclusive and equitable learning opportunities. By addressing the initial challenges of high costs and development time, the construction industry and academia can fully leverage VR's potential in construction practices and education.

## 6. Conclusion

This systematic review thoroughly examined the impact of virtual reality (VR) applications on construction education, training, and practice. The findings reveal a wide range of positive outcomes in these areas. The collected evidence indicates that VR improves learning outcomes by providing immersive and interactive experiences while also enhancing the safety and efficiency of construction practices.

VR's ability to simulate complex construction processes in a controlled environment represents a significant advancement over traditional learning and training methods. This enables students to visualize and interact with construction project elements and sequences in ways that deepen their understanding and retention. As the construction industry continues to advance, the need for innovative educational tools is expected to grow, positioning VR as a valuable asset in preparing a skilled workforce.

To fully harness the benefits of VR in construction education, construction educators should consider adopting a hybrid approach that integrates traditional teaching methods with VR technology. This approach aims to offer a well-rounded educational experience that capitalizes on the strengths of both methods while minimizing their limitations. In the future, researchers should concentrate on developing cost-effective VR solutions and exploring how VR can be effectively combined with other emerging technologies, such as AR and MR. These developments have the potential to enrich the educational environment by providing more accessible and adaptable learning tools, as seen in this review.

Although VR holds significant promise for transforming construction education and work practices, its seamless integration into mainstream applications requires deliberate planning, ongoing technological progress, and insights driven by research. By addressing these challenges and effectively leveraging the capabilities of VR, the construction industry can meet present and future demands, ultimately preparing a new generation of professionals with the knowledge and skills needed to excel in a technologically advanced industry.

**Declaration of generative AI and AI-assisted technologies in the writing process**
During the preparation of this work, the author(s) used ChatGPT and Grammarly to assist with refining the language and structure. After using these tools/services, the authors reviewed and edited the content as needed and take full responsibility for the content of the publication.

**Data Availability Statement:** The data that support the findings of this study are included in the paper.